\documentclass[aps,prl,reprint,showpacs,showkeys]{revtex4-1}
\usepackage{microtype}
\usepackage{amsthm}
\usepackage{amsmath}
\usepackage{amsfonts}
\usepackage[mathscr]{eucal}
\usepackage{amssymb,epsfig,setspace}
\usepackage{graphicx}
\theoremstyle{remark}

\usepackage{natbib}
\usepackage[usenames]{xcolor}
\usepackage[textwidth=4cm,textsize=small,shadow]{todonotes}

\begin{document}
%
\title[An exactly solvable system from quantum optics]{
An exactly solvable system from quantum optics}
\author{Andrzej J.~Maciejewski} \email{maciejka@astro.ia.uz.zgora.pl}
\affiliation{J.~Kepler Institute of Astronomy, University of Zielona
  G\'ora, Licealna 9, PL-65--417 Zielona G\'ora, Poland.}%
\author{Maria Przybylska}%
\email{M.Przybylska@if.uz.zgora.pl} \affiliation{ Institute of
  Physics, University of Zielona G\'ora, Licealna 9, 65--417 Zielona
  G\'ora, Poland }%
\author{Tomasz Stachowiak} \email{stachowiak@cft.edu.pl}
\affiliation{%
  Center for Theoretical Physics PAS, Al. Lotnik\'ow 32/46, 02-668
  Warsaw, Poland }%

\date{\today}%
\begin{abstract}
  We investigate a generalisation of the Rabi system in the Bargmann-Fock
  representation.  In this representation the eigenproblem
  of the  considered quantum model is described by a system of two linear
  differential equations with one independent variable. The system has
  only one irregular singular point at infinity. We show how the quantisation
  of the model is related to asymptotic behaviour of solutions in a
  vicinity of this point.  The  explicit formulae for
  the spectrum  and eigenfunctions of the model follow from
  an analysis of the Stokes phenomenon.  An interpretation of the obtained
  results in terms of differential Galois group of the system is also
  given.
\end{abstract}
                      
\maketitle
\section{Introduction and results}
In paper~\cite{Maciejewski:14::a} we proposed a general method which
allows to determine spectra of quantum systems given in the
Bargmann representation. This representation is very useful for
systems with a Hilbert space which is a product of a finite and
infinite-dimensional spaces. A typical example is a
Hamiltonian for spin degrees of freedom of some particles, characterised by
$n$-dimensional $\sigma$ matrices, coupled to a bosonic field via annihilation
and creation operators $a$, and $a^{\dag}$. Then $n$-component wave
function $\psi=(\psi_1,\ldots, \psi_n)$ is an element of Hilbert space
$\mathscr{H}^n=\mathscr{H}\times\cdots\times \mathscr{H}$, where
$\mathscr{H}$ is the Bargmann-Fock Hilbert space of entire functions.
The scalar product in $\mathscr{H}$ is given by
\[
\langle
f,g\rangle=\dfrac{1}{\pi}\int_\mathbb{C}\overline{f(z)}g(z)\mathrm{e}^{-|z|^2}
\,\mathrm{d} x\,\mathrm{d}y, \quad z=x+\mathrm{i}\, y.
\]
Since operators $a$ and $a^{\dag}$ are represented by $\partial_z$ and
multiplication by $z$, respectively, for clearly $[\partial_z,z]=1$,
stationary Schr\"odinger equation $H\psi=E\psi$ becomes the system
of linear ordinary differential equations. Usually such systems have a
certain number of regular singular points in the complex $z$ plane,
and possibly an irregular point at infinity.  Determination of the
spectrum consists in finding such values of $E$  that all the
components of wave function $\psi_i$, with $i=1,\ldots,n$, are
elements of the Hilbert space $\mathscr{H}$.

In paper~\cite{Maciejewski:14::a} we have shown that a solution of the
eigenvalue problem can be reduced to checking the following three
conditions.
\begin{enumerate}
\item Local conditions. At each regular singular point $z=s$ there
  exists at least one solution which is holomorphic in an open set
  containing $s$.
\item Global conditions. At each singular point we can select a local
  holomorphic solution in such a way that they are analytic
  continuations of each other. That is, they are local representations
  of an entire solution.
\item Normalisation conditions. The entire solution selected above 
  must have a finite Bargmann norm.
\end{enumerate}

The first two conditions were analysed in detail
in~\cite{Maciejewski:14::a} and in \cite{Maciejewski:14::} we gave
their effective application for determination of full spectrum of the
Rabi system.

The third condition is highly non-trivial. This is so because ``most''
of entire functions do not belong to  $\mathscr{H}$.  In order to
analyse  this condition we need to characterise the growth
of an entire solution $f(z)$ of a system of linear differential equations.  
In a neighbourhood of a regular singular
point the growth of $f(z)$ is polynomial. Hence, if the considered
system has only regular singular points, and $f(z)$ is its entire
solution, then $f(z)$ has a finite Bargmann norm.  Normalisation
conditions can give non-trivial constrains  on entire solution only if the linear system
has an irregular singularity, for example at infinity.

The growth of entire function $f(z)$ is described by means of the
following function:
\begin{equation}
  M(r) := \underset{|z|=r}{\mathrm{max}}|f(z)|.
\end{equation}
It is used to define two numbers which characterise properties of the
growth. The order, or the growth order $\varrho$ of $f(z)$ is defined
as the limit
\begin{equation}
  \varrho:=\lim_{r\rightarrow\infty}\sup\frac{\ln(\ln M(r))}{\ln r} 
  ,\quad
  \mathrm{with}\quad 0\leq \varrho \leq \infty.
\end{equation}
For an entire function of finite order $\varrho<\infty$, its type
$\sigma$ is defined as
\begin{equation}
  \sigma:=\lim_{r\rightarrow\infty}\sup\frac{\ln M(r)}{r^\varrho}.
\end{equation}
If $f(z)$ belongs to $\mathscr{H}$, then one can prove the following
facts \cite{Bargmann:61::}:
\begin{enumerate}
\item $f(z)$ is of order $\varrho\leq 2$.
\item If $\varrho=2$, then $f(z)$ is of type $\sigma\leq\tfrac12$.
\end{enumerate}
If $\varrho=2$ and $\sigma=\tfrac12$, then the question whether
$f(z)\in\mathscr{H}$ requires a separate investigation.  For
additional details see \cite{Vourdas:06::}.

It is well known that in a neighbourhood of a regular singular point,
a formal procedure allows to find formal series which satisfy the
equation. It appears that this formal series is convergent, so it is 
a local solution of the equation.  In a
vicinity of irregular singular point a formal procedure gives formal
expressions which satisfy the equation, however the series are
generally divergent. It is known that these formal expressions, give
the asymptotic expansion of a solution valid in in certain sectors
radially extending from the vertex localised at this singularity.
Borders of sectors are determined by the so-called Stokes lines. The
asymptotic expansions change when we pass from one sector to
another.  These changes are governed by the so-called Stokes matrices.

The normalisation condition of an entire solution $f(z)$ implies that
at each sector $S$ the following integral 
\begin{equation}
  \label{eq:1}
  \int_S |f(z)|^2 \mathrm{e}^{-|z|^2} \, \mathrm{d} x \, \mathrm{d} y,
\end{equation}
has a finite value.  Thus, at each sector it has good asymptotics
which guarantees the convergence of the above integral.  This implies
that although a solution $f(z)$ can change its asymptotic expansion from
sector to sector, these changes are restricted only to good
asymptotics.  This property implies that all Stokes matrices have a
common invariant subspace, in particular case just one common
eigenvector.

The aim of this paper is to show that the above considerations can be
applied effectively.  We consider system given by the following
Hamiltonian
\begin{equation}
  H= \left(\omega +\dfrac{U}{2}\sigma_z\right) a^{\dag}a + \dfrac{\omega_0}{2}\sigma_z + 
  g\sigma_x(a^{\dag}+a),
  \label{hamster}
\end{equation}
where $\sigma_x$, $\sigma_z$ are the Pauli spin matrices and
$\omega_0$, $\omega$, $g$ and $U$ are parameters. For $U=0$, it
coincides with the Hamiltonian of the Rabi model \cite{Rabi:36::}
describing interaction of a two-level atom with a single harmonic mode
of the electromagnetic field. Hamiltonian \eqref{hamster} was proposed
in \cite{Grimsmo:13::a,Grimsmo:14::a}.  The term $\frac{U}{2}\sigma_z
a^{\dag}a$ can be interpreted as a nonlinear coupling between the atom
and the cavity.

In Bargmann-Fock representation, the stationary Schr\"odinger equation
$H\psi=E\psi$, with $\psi=(\psi_1,\psi_2)$, have the form
\begin{equation}
  \begin{split}
    \left(\omega+\dfrac{U}{2}\right)z\psi_1'+
\dfrac{\omega_0}{2}\psi_1+g\psi_2'+gz\psi_2=E\psi_1,\\
    \left(\omega-\dfrac{U}{2}\right)z\psi_2'-
 \dfrac{\omega_0}{2}\psi_2+g\psi_1'+gz\psi_1=E\psi_2.
  \end{split}
  \label{eq:bargmanki}
\end{equation}
In our paper \cite{Maciejewski:14::a} we calculated the spectrum
of $H$ for the generic case, i.e., when the above system has two
regular singular points. This requires that $U^2\neq4\omega^2$, and
the quantisation of the energy spectrum results from the condition that
system~\eqref{eq:bargmanki} admits an entire solution. In that case,
all entire solutions of~\eqref{eq:bargmanki} have a finite norm.

In this paper we investigate the remaining cases for which
$U^2=4\omega^2$. For $U=+ 2 \omega$ system~\eqref{eq:bargmanki} can be
rewritten in the form
\begin{equation}
  \begin{split}
    \psi_1'(z) =& -z \psi_1(z) +\frac{2E+\omega_0}{2g}\psi_2(z),\\
    \psi_2'(z) =& \frac{4\omega z^2 +2E-\omega_0}{2g}\psi_1(z)
    - \\
    & \frac{z\left[g^2+\omega(2E+\omega_0)\right]}{g^2}\psi_2(z).
  \end{split} \label{eq2}
\end{equation}
The system for the case of $U=-2\omega$ can be obtained from the above
equations by a simple change $\omega_0\rightarrow -\omega_0$, and the
interchange  $\psi_1$ with $\psi_2$.  Hence, we consider only the case
$U=+ 2 \omega$.

System~\eqref{eq2} has no singular points in the finite part
of $\mathbb{C}$, so all its solutions are entire functions. Infinity
is the only and irregular singular point.  The system can have
solutions which grow fast enough to make the Bargmann norm infinite.
Thus, to determine the spectrum of the problem we have to find
all  values of $E$ for which system  admits a solution
with a moderate  growth at infinity, such that  its  Bargmann
norm is finite.

We give a full answer to this question. That is we specify explicitly
a countable number of energy values for which the corresponding
eigenfunctions are also given explicitly and have a finite Bargmann
norm.

 The energy axis is divided into two disjoint intervals, each
of them with its own countable family of eigenvalues.  To be more
precise, our main result is as follows. Let
\begin{equation}
  \label{eq:2}
  x:=1+\omega
  g^{-2}(E+\omega_0/2),
\end{equation}
be an auxiliary spectral parameter.  Then the Hamiltonian
\eqref{hamster} has entire solutions with a finite Bargmann norm
if and only if one of the following conditions is fulfilled. 
Either $x>1$, and
\begin{equation}
\label{w+}
\frac{(x-1)\left(\omega(\omega-\omega_0)+g^2(x-1)\right)}
  {\omega^2\sqrt{x^2-1}} = 2n+1,
\end{equation}
or $x<-1$, and
\begin{equation}
\label{w-}
  \frac{(x-1)\left(\omega(\omega-\omega_0)+g^2(x-1)\right)}
  {\omega^2\sqrt{x^2-1}} = -(2n+1),
\end{equation}
where $ n\in\mathbb{N}$.  The respective eigenfunctions
\[
\psi^{\pm}_n=\left(\psi_{1,n}^{\pm},\psi_{2,n}^{\pm}\right),
\]
are given by
\begin{equation}
  \begin{split}
    \psi_{1,n}^{\pm}(z) &= C \exp\left(-\beta_{\pm}z^2\right)
    H_{n}\left(\sqrt{\pm1}\sqrt[4]{x^2-1}\, z\right),\\
    \psi_{2,n}^{\pm}(z) &=
    \frac{\omega}{g(x-1)}\left[\psi_{1,n}^{\pm}(z)+
      \left(\psi_{1,n}^{\pm}(z)\right)'\right],
  \end{split}
\end{equation}
where $H_n(z)$ denotes the Hermite polynomial of degree $n$, and
\begin{equation*}
  \beta_{\pm}:= \frac{x\pm\sqrt{x^2-1}}{2}.
\end{equation*}
These functions have the growth order $\varrho=2$ and type
$\sigma=\beta_{\pm}$.

\section{Asymptotic expansions}

An asymptotic expansion of a function $\psi(z)$ for $z\rightarrow
\infty$ is denoted in the following way
\begin{equation}
  \psi(z)\sim a_0 + \frac{a_1}{z} + \frac{a_2}{z^2} +\ldots\,. 
\end{equation}
By definition it means  that in a given sector 
\begin{equation*}
S(a,b):= \{ z\in\mathbb{C} \, |\,  a<\arg(z)< b, \quad |z|>r_0\},
\end{equation*}
we have 
\begin{equation}
  \lim_{z\rightarrow\infty} 
  \left(\psi(z) - \sum_{j=0}^{m}a_j z^{-j} \right)z^{m} = 0,
\end{equation}
for an arbitrary $m\in\mathbb{N}$, and $z\in S(a,b)$. Such series needs not
have a positive radius of convergence.

Note that if 
\begin{equation}
  \psi(z) e^{-P(z)} \sim a_0+\mathcal{O}(z^{-1}),
\end{equation}
for a certain function $P(z)$, then,  in particular 
\begin{equation}
 \lim_{z\rightarrow\infty} \psi(z)e^{-P(z)}= a_0,
\end{equation}
 and $  |\psi(z)| =\mathcal{O}(\exp |P(z)|)$.

The general form of solutions around irregular singular points
is known, and there are several classical theorems on them. We choose
two, particularly suitable here, namely Theorems 12.3 and 19.1 given in
 \cite{Wasow:87::}, which tell as the following.

Let us suppose that the system in question can be written as
\begin{equation}
  z^{-q} \psi'(z) = A(z)\psi(z), \qquad \psi(z) \in\mathbb{C}^n,
  \label{standform}
\end{equation}
where $A(z)$ is $n\times n$ matrix  holomorphic in a neighbourhood of infinity and
$A_0:=A(\infty)$ is nonzero.  Then in every sufficiently narrow sector
$S$, the equation has a fundamental matrix of solutions of the form
\begin{equation}
  {\Psi}(z) = Y(z) z^G e^{Q(z)},
\end{equation}
where $G$ is a constant $n\times n$ matrix, $Q(z)$ is a diagonal matrix whose
elements are polynomials of $z^{1/p}$ with a suitable positive integer
$p$, and $Y(z)$ admits an asymptotic power series
in $z^{-1/p}$. If we further assume that the eigenvalues  $\lambda_i$
of $A_0$, are all distinct, then 
\begin{itemize}
\item $G$ is a diagonal matrix,
\item $p=1$ and $Q(z)$ is also diagonal with leading terms
  of $\lambda_i z^{q+1}/(q+1)$,
\item $S$ can be any sector of central angle not exceeding
  $\pi/(q+1)$.
\end{itemize}
The fact that asymptotic expansions are unique allows for their
formal calculation, and the above tells us, that there are always
solutions which grow according to the prescribed formal
expressions. The problem is that for a true (not just formal) solution the
asymptotic
behaviour  changes from sector to sector---only in rare cases are the
expansions valid for all values of the argument. 

It can be shown, see \cite{Wasow:87::}, that if a
function is single valued near infinity, its asymptotic expansion
is valid for all values of $\arg(z)$ if and only if it is analytic
at $z=\infty$ and the series is actually convergent .

In our case the requirement of finite Bargmann norm is  that
\begin{equation}
  |\psi(z)|= \mathcal{O}\left( e^{\sigma|z|^{\varrho}}\right),\quad \varrho\leq2,
\end{equation}
and $|\sigma|\leq\tfrac12$ when $\varrho=2$. So the determination
of the exponential factor will be crucial, and we shall see that the
considered model presents us with functions of growth order
$\varrho=2$ and varying types $\sigma$, which require detailed
analysis.

\section{ Spectrum determination}

In the  case $U=+2\omega$ the considered model is described by~\eqref{eq2}. 
This system has  the standard  form \eqref{standform} with
$q=2$. Since  the leading matrix $A_0$ is nilpotent, 
it is convenient to perform a shearing transform
\begin{equation}
  \psi_{i}(z) = z^{i-1}\xi_i(z), \quad i=1, 2,
\end{equation}
after which the system becomes
\begin{equation}
  \frac{1}{z} \boldsymbol{\xi}'(z) =
  \begin{bmatrix}
    -1 & (x-1)\frac{g}{\omega} \\
    \frac{2\omega}{g} +\frac{g^2(x-1)-\omega\omega_0}{g\omega z^2} &
    1-2x -\frac{1}{z^2}
  \end{bmatrix}
  \boldsymbol{\xi}(z),
\end{equation}
where $\boldsymbol{\xi}= [\xi_1, \xi_2]^T$, $x$ is a new spectral
parameter defined by~\eqref{eq:2}.  The rank of the system is thus
$q+1=2$, and the matrix $A_0$ has eigenvalues
$x\pm\sqrt{x^2-1}$. We can immediately discard the case of them being
equal, for then, the system \eqref{eq2} is explicitly solvable. For
example, when $x=1$ we have
\begin{equation*}
  \psi_1(z) = c_1 e^{-z^2/2},\,\,
  \psi_2(z) = \frac{c_2}{3g}e^{-z^2/2}(2\omega z^3-3\omega_0 z),
\end{equation*}
and these functions have  infinite Bargmann norm.

We are thus left only with the simpler case of two distinct
eigenvalues and we know what form of asymptotic expansions to expect.
Following Birkhoff \cite{Birkhoff:1909::}, let us assume the following
Laplace integral representation for the vector of solutions
\begin{equation}
  \psi_i(z) = \frac{1}{2\pi \mathrm{i}}\int_{C} 
  \exp(z^2 y)(\eta_{0,i}(y)+z\eta_{1,i}(y))\mathrm{d}y,
  \label{lap}
\end{equation}
where $i=1,2$.
The usual inverse Laplace transform would be
\begin{equation}
  \frac{1}{2\pi \mathrm{i}}\int_{C} \exp(z y)\eta(y)\text{d}y,
\end{equation}
However, such $\eta$ would, in general, have essential singularities
in the finite complex plane because we are dealing with an equation
of rank 2. To remedy this, the exponential part needs to be modified
in anticipation of the growth order 2. This procedure is applicable
whenever the matrix $A_0$ has different eigenvalues, regardless of the
rank of the equation, for further details see~\cite{Birkhoff:1909::}.

\begin{figure}[ht]
  \begin{center}
    \includegraphics[width=.9\columnwidth]{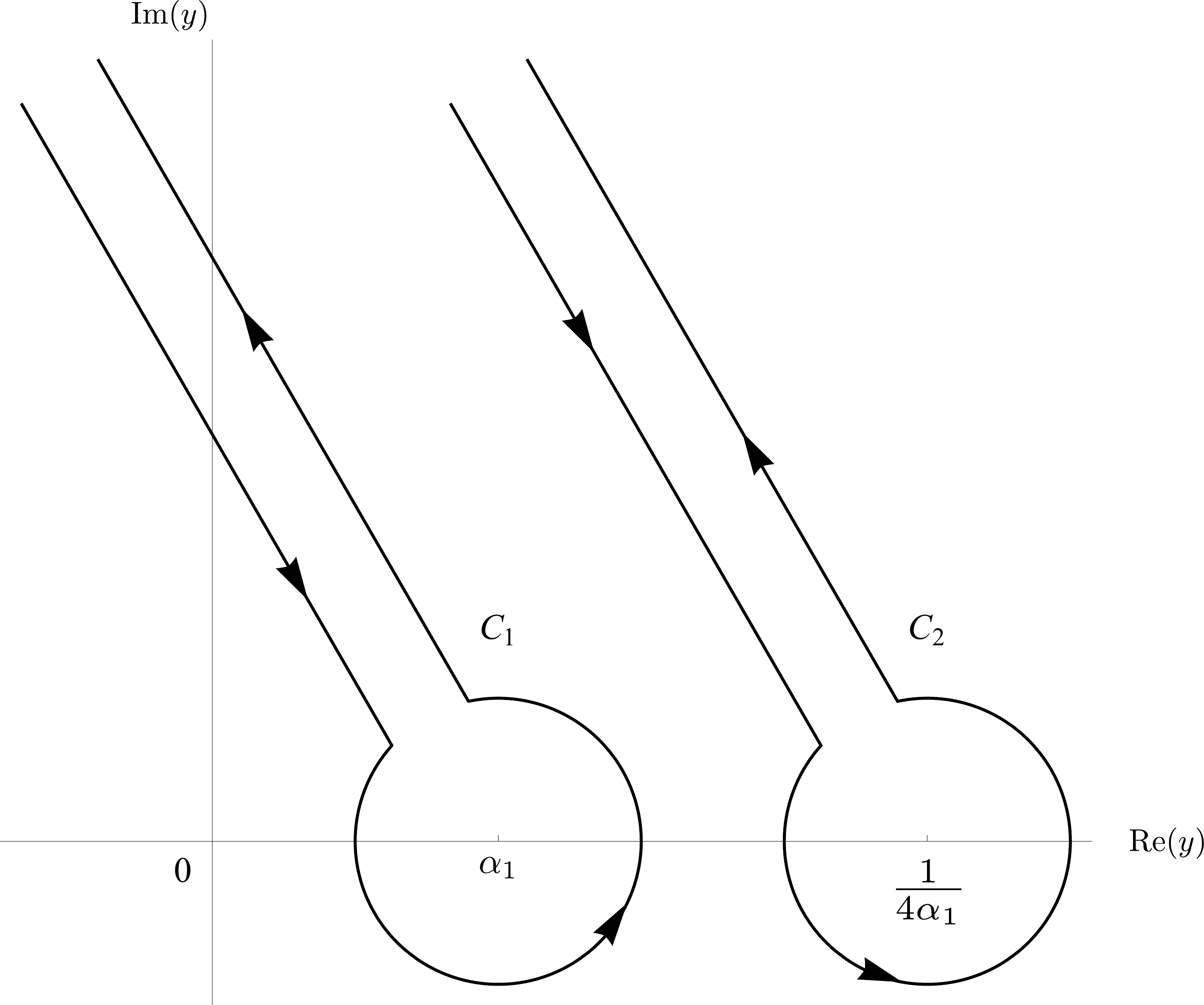}
    \caption{Contours. $\alpha_1=\delta=1/(4\alpha_2)$}
    \label{fig1}
  \end{center}
\end{figure}

The contour of integration $C$ is chosen so that the integrand assumes the
same value at both ends. It could be a loop or an open contour extending to
infinity. Here, the latter is more viable, and $C$ will be one of the curves
shown in Figure \ref{fig1}.  This will allow to integrate by parts as follows
\begin{multline}
  \int_C \exp(z^2 y)z^n u(y)\mathrm{d}y =\\ \left[\exp(z^2
    y)z^{n-2}u(y)\right]_C -\int_C\exp(z^2 y)z^{n-2}u'(y)\mathrm{d}y,
  \label{parts}
\end{multline}
and taking the bracket term to be zero. Substituting the Laplace
representation \eqref{lap} in equation \eqref{eq2} we obtain a new
system for $(\eta_{0,i},\eta_{1,i})$ with $i=1,2$, by equating the
integrands to zero. That this is permissible can be checked later
where the particular $\eta$ are recovered but we are mostly interested
in the asymptotic expansions in which case the expressions obtained
will be formal anyway.

The integrands are  polynomials of $z$ and integrating by parts
as in \eqref{parts} we can reduce their degree to at most 1. Then the
coefficients of $z^0$ and $z^1$ must each be zero so that we end
up with a system of four equations
\begin{equation}
  \begin{split}
    \omega \eta_{11}+g(x-1)\eta_{02}+\omega(2y+1)\eta_{11}' &= 0,\\
    \omega(2y+1)\eta_{01}-g(x-1)\eta_{12} &= 0,\\
    \omega(2\omega\eta_{01}'-g(2y+2x-1)\eta_{12}'-g\eta_{12}) &\\
    + (\omega_0\omega-g^2(x-1))\eta_{01}&= 0,\\
    (\omega_0\omega-g^2(x-1))\eta_{11} + & \\
    \omega(2\omega\eta_{11}'+g(2y+2x-1)\eta_{02}) &= 0,
  \end{split}
\end{equation}
where prime denotes differentiation with respect to $y$.
First two equations can be  immediately  solved
for two components
\begin{equation}
  \begin{split}
    \eta_{02} &=
    -\frac{\omega}{g(x-1)}\left((2y+1)\eta_{11}'+\eta_{11}\right),\\
    \eta_{12} &= \frac{\omega(2y+1)}{g(x-1)}\eta_{01}.
  \end{split}
\end{equation}
This leaves  two first order equations for the remaining components
\begin{equation}
  \begin{split}
    (4y^2+4xy+1)\eta_{10}' &= \left(4\rho\sqrt{x^2-1}-6y-3x\right)\eta_{10},\\
    (4y^2+4xy+1)\eta_{11}' &= \left(4\rho\sqrt{x^2-1}-2y-x\right)\eta_{11},\\
    \label{decoup}
  \end{split}
\end{equation}
where
\begin{equation}
  \rho = -\frac{(x-1)\left(\omega(\omega-\omega_0)+g^2(x-1)\right)}
  {4\omega^2\sqrt{x^2-1}}.
\end{equation}
Although the particular equations decouple, this is obviously not a
generic feature. It makes sense then to treat them as a system with
two regular singular points at
\begin{equation}
  \alpha_{1,2} = \frac12\left(-x\pm\sqrt{x^2-1}\right),
\end{equation}
and one at infinity. The characteristic exponents are
\begin{equation}
  \begin{split}
    \beta_1\in\left\{\rho-\tfrac34,\rho-\tfrac14\right\}
    \quad &\text{at } \alpha_1,\\
    \beta_2\in\left\{-\rho-\tfrac34,-\rho-\tfrac14\right\}
    \quad &\text{at } \alpha_2,\\
    \beta_{\infty}\in\left\{\tfrac32,\tfrac12\right\} \quad
    &\text{at } \infty.
  \end{split}
\end{equation}

In general, the behaviour of a solution as it is continued around all
the singular points cannot be specified explicitly as the monodromy
group of the equation is unknown. Fortunately, solutions
of system~\eqref{decoup} are  given explicitly
\begin{equation}
  \begin{split}
    \eta_{01} &= b_1 (y-\alpha_1)^{\rho-3/4}(y-\alpha_2)^{-\rho -3/4},  
    \\
    \eta_{11} &= b_2 (y-\alpha_1)^{\rho-1/4}(y-\alpha_2)^{-\rho -1/4},
  \end{split}
  \label{lapeq}
\end{equation}
where $b_{1,2}$ are arbitrary constants. The explicit forms 
of these solutions  will give us full
information about which contours to choose and how the continuation
of any solution behaves.

Each singular point of the solution $\eta_{i,j}(y)$ 
corresponds to a different
exponential factor in the asymptotic expansion, which can be seen
as follows.  Let us look at the generic case first, when the contour
can be chosen as path $C_i$ that runs from infinity to the vicinity
of the point $y=\alpha_i$, around it in the positive direction, then
back to infinity, as depicted in Figure~\ref{fig1}.
This requires that the characteristic exponent at that singularity $\alpha_i$
is not an integer, for otherwise the integral vanishes.   
Since, for the system~\eqref{decoup}   for all singular points 
the differences of  exponents is  $1/2$, one solution is always
suitable.  

The Laplace integral \eqref{lap} for any function $\eta(y)$ which has a
local power series around $y=\alpha$ with characteristic exponent
$\beta$ may be rewritten as
\begin{multline}
  \frac{e^{\alpha z^2}}{2\pi \mathrm{i}}\int_{C}
  e^{z^2(y-\alpha)}(y-\alpha)^{\beta}
  \sum_{n=0}^{\infty}a_n(y-\alpha)^n\mathrm{d}y\\= \frac{e^{\alpha
      z^2}}{z^{2(\beta+1)}}\sum_{n=0}^{\infty}
  \frac{a_n}{\Gamma(-\beta-n)z^{2n}},
\end{multline}
where we have used the integral representation of the gamma function,
and omitted indices for brevity. The series is not, in general,
convergent in the whole complex plane of $y$, so the integration term
by term using the gamma function leads to a formal, divergent in most
cases, asymptotic expansion. It can be shown that this is the correct
asymptotic series \cite{Kohno:99::}, so that it is immediately evident
that each singular point $\alpha_i$ gives rise to one type
of exponential factor $\exp(\alpha_i z^2)$ in the asymptotic series. Note that
$4\alpha_1\alpha_2 = 1$ and in view of the previous section we require
$|\alpha_i|<1/2$, so only one of the contours will lead to the desired
growth type. The special case of $|\alpha_i|=1/2$ will be dealt with
separately, below.

\begin{figure*}[ht]
  \begin{center}
    \includegraphics[width=.49\textwidth]{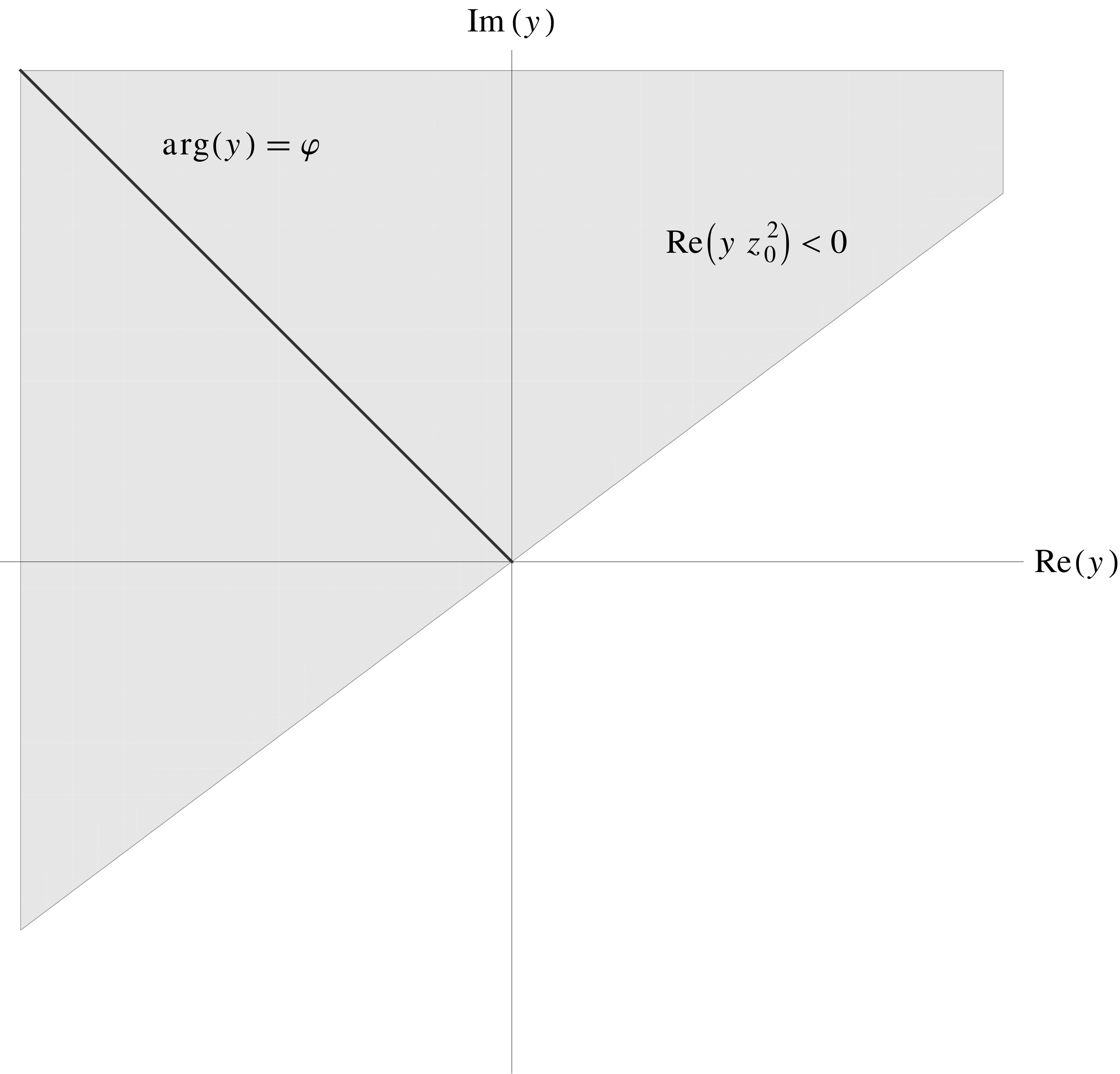}
    \includegraphics[width=.49\textwidth]{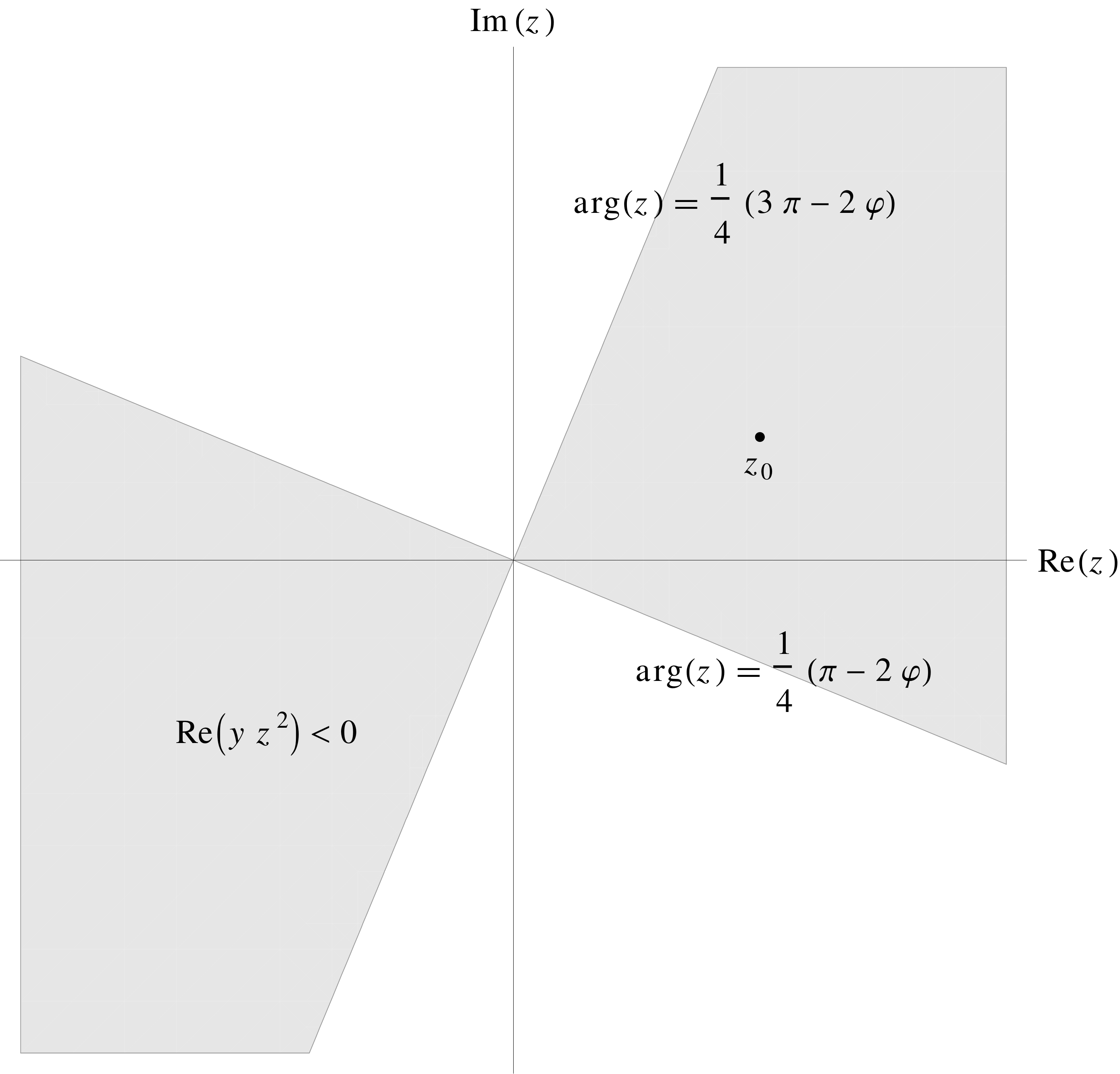}
    \caption{The corresponding sectors in the $y$ and $z$ planes for
      $z_0=1+\mathrm{i}/2$. Note that both shaded regions on the right fulfil
  $\text{Re}(yz^2)<0$ for any given $y$ with $\arg(y)=\varphi$.}
    \label{fig2}
  \end{center}
\end{figure*}

It is important to keep in mind that the contour had to be chosen so
that the integrand vanishes as $y\rightarrow\infty$ and that the
integral \eqref{lap} is convergent -- this is ensured by
$\text{Re}(z^2y)<0$. In other words, for a given argument of $z$ there
is just a sector of the complex plane of $y$ where the integral
converges and vice versa: for a given ray ($\arg(y-\alpha)$) in the
$y$ plane, there is a region of convergence in the
$z$ plane. An example is shown in Figure \ref{fig2}. Because the
components $\eta$ grow at most with some power of $y$ at infinity, the
integral \eqref{lap} converges fast enough, for the function to be
holomorphic and it is indeed permissible to differentiate
it as we have done.

\begin{figure}[ht]
  \begin{center}
    \includegraphics[width=.9\columnwidth]{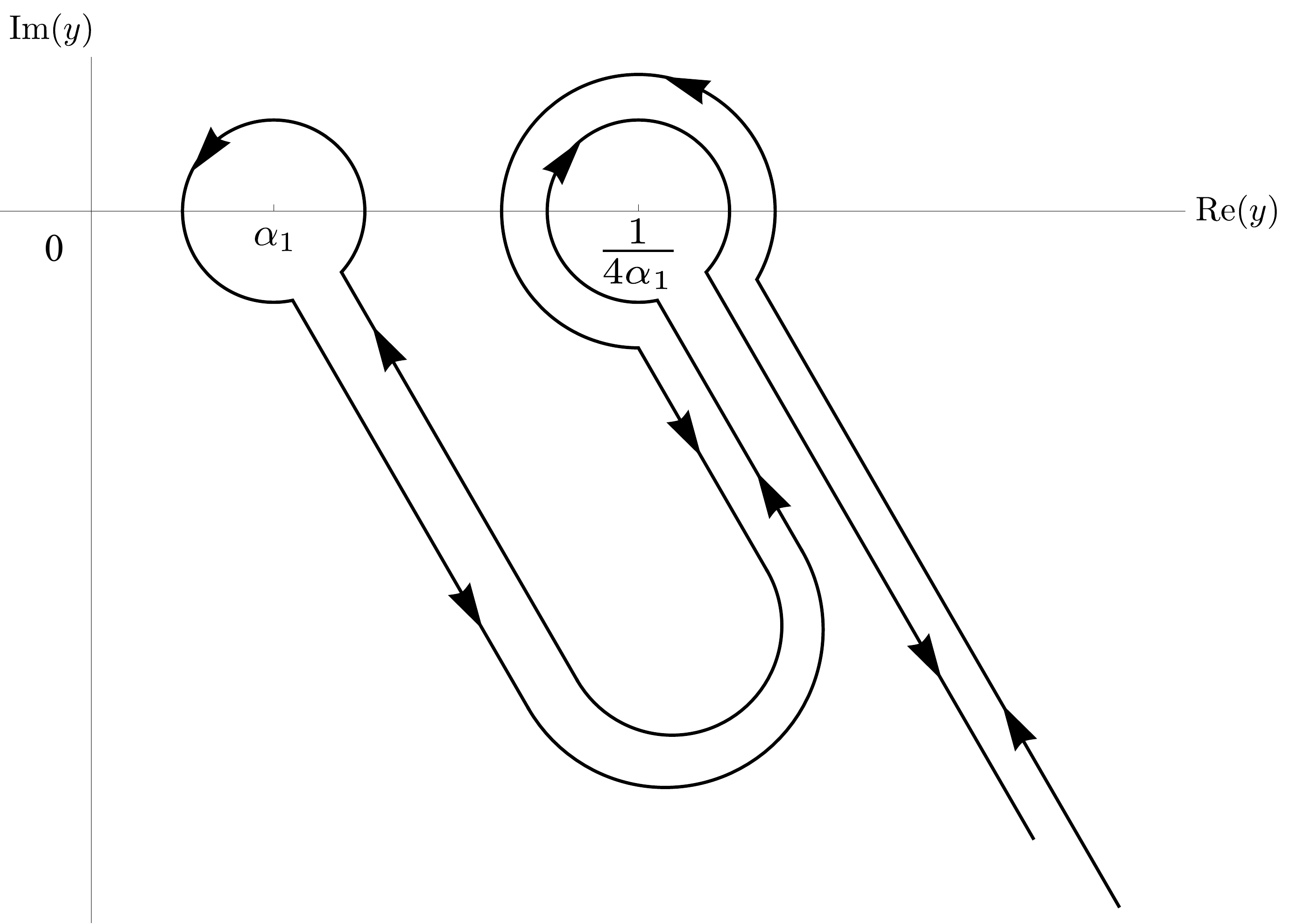}
    \caption{Deformation.}
    \label{fig3}
  \end{center}
\end{figure}

To extend the function outside one sector, the integral can be
analytically continued by deforming the contour, rotating it around a
given singularity.  Because the contour has to pass through the other
singularity, the solution associated with one asymptotics could
transform into a combination of both, as shown in figure
\ref{fig3}. This is the Stokes phenomenon, and the argument
of $\alpha_1-\alpha_2$ determines the so-called Stokes lines in the
$z$ plane, which is customarily given by the condition
$\text{Re}(z^2(\alpha_1-\alpha_2))=0$; crossing the line causes the
asymptotic expansions to mix. There is the freedom of choosing the
starting contour, which corresponds to choosing a particular solution,
and the first Stokes line where the deformation is required depends
on that choice. But if we continue a full basis increasing the
argument of $z$ we eventually need to take into account all of the
lines. For the considered case, $\alpha_1-\alpha_2$ is real, so there are two Stokes lines 
which pass through the origin and form angles $\pm \pi/4$ with the $x$-axis.  


Upon crossing, the original integral decomposes into integrals over
new contours, which could be written shortly as
\begin{equation}
    \mathcal{A}\int_{C_i} = \int_{C'_j}+\,
  \int_{C'_i}  
  -\,e^{2\pi\mathrm{i}\beta_i}\int_{C'_j},
\end{equation}
where $\mathcal{A}$ stands for analytic continuation, and primes are used to
distinguish between contours in the two
regions: $\text{Im}(y)>0$ and $\text{Im}(y)<0$, although their shape remains
the same. This could be written
using the Stokes matrix as
\begin{equation}
    \mathcal{A}
  (e_{i},e_{j}) = 
  (e'_{i},e'_{j})
  \begin{pmatrix}
    1 & 0\\
    1-e^{2\pi \mathrm{i}\beta_i} & 1
  \end{pmatrix},
\end{equation}
where $e_i$ is the solution obtained with $C_i$, hence with fixed asymptotics
of $\exp(\alpha_i z^2)$, and likewise for $e'_i$ and $C'_i$. By convention, the
matrix on the right-hand side is the inverse of the Stokes matrix.

As $\beta_i$ is not
an integer, whether the additional
integral actually contributes, depends on the characteristic exponent
of the solution around $\alpha_j$. If $\beta_j$ is not an integer
either, the Stokes multiplier is obviously not zero. On the other
hand, for an integer $\beta_j$ the contour $C_j$ can simply be
deformed through the singularity so the integral is identically zero,
because
\begin{equation}
  \int_{C_j} 
  = \left(1-e^{2\pi \mathrm{i}\beta_j}\right)\int_{\alpha_j}^{\infty}
  =: \left(1-e^{2\pi \mathrm{i}\beta_j}\right) \int_{L_j}.
\end{equation}
This does not mean that the asymptotic solution $e_j$ vanishes, but
simply that the contour $L_j$, which is a half line from $\alpha_j$ to
infinity, has to be used instead of $C_j$ to define the other
asymptotic solution $e_j$.  The Stokes matrix is then
\begin{equation}
    \mathcal{A}(e_{i},e_{j}) = (e'_{i},e'_{j})
  \begin{pmatrix}
    1 & 0\\
    0 & 1
  \end{pmatrix}.
\end{equation}
For completeness we note that the contour $L_j$ can be deformed
analogously to the above, this time winding around $\alpha_i$. Because
$\beta_i$ is not an integer, the contour cannot pass through the
singular point and the additional path will be exactly $C_i$. In terms
of the Stokes matrix we have
\begin{equation}
    \mathcal{A}(e'_{i},e'_{j}) = (e''_{i},e''_{j})
  \begin{pmatrix}
    1 & 1\\
    0 & 1
  \end{pmatrix}.
\end{equation}
In other words, the asymptotics of the solution $e_i$ remains
unchanged, while the asymptotic series $e_j$ actually exhibits the
Stokes phenomenon because $\beta_i$ is not an integer. The double primes
signify contours identical to the original ones for $\text{Im}(y)>0$ but the
solutions are not single-valued in general so $e_i''$ are some multiples of
$e_i$.

Note, that when $\beta_j$ is an integer, $\beta_i$ cannot be, since
their sum is $-3/2$ or $-1/2$ for the two solutions
of $\eqref{lapeq}$, respectively.  Thus, the conditions that the
contour is suited for the noninteger exponent at $\alpha_i$ and that
the Stokes multiplier vanishes are consistent. There is no need then
to investigate the case of integer $\beta_i$, because for any $\rho$
we always have at least one solution with noninteger exponent at each
of the singular points. It follows that we always have two different solutions
$e_i$ with different asymptotic expansions obtained by means of the contours
$C$ and,
as stated in the previous section, the fundamental solutions have to
be expressible by means of those two $e_i$ in any sector. In each case
then, we know how both asymptotic solutions mix when crossing the
Stokes lines and hence, how the growth of any actual solution
changes. For the system under consideration the situation is very
simple, because only the asymptotic series with smaller
$\alpha_i$ is suitable, so that all of its Stokes multipliers must be
zero to ensure finite norm.

It remains to verify that when $x^2<1$ and hence $|\alpha_i|=1/2$ the
resulting solutions are not normalisable, in close analogy to the
$x=1$ case, where the components of the wave function behave like
$e^{z^2/2}$. The exponential factor in the present case can be written
as
\begin{equation}
  \exp\left(\frac12 e^{\mathrm{i}\theta} z^2\right),\quad\text{for }
  \alpha_j = \frac12 e^{\mathrm{i}\theta_j},
\end{equation}
and since the equation determining $\alpha_j$ is real we have
$\alpha_2=\bar{\alpha_1}$ so $\theta_2=-\theta_1\neq 0$.  Together
with the Bargmann measure this results in the leading factor of the
form
\begin{equation}
  \int_0^{2\pi}\int_0^{\infty} 
  \exp\left(-2r^2\sin^2\left(\varphi+\frac{\theta}{2}\right)\right)
  r\text{d}r\text{d}\varphi,
\end{equation}
where we took $z=r e^{\mathrm{i}\varphi}$. The integral over $r$ converges
provided that the sine does not vanish, i.e. in sectors that do not
include the directions $\varphi+\theta_j/2=k\pi$,
$k\in\mathbb{Z}$. If there is no Stokes phenomenon for the solutions
with asymptotic factors of $e^{\alpha_i z^2}$, it follows that the
norm is not finite. If the two solutions could interchange, we would
necessarily have to construct a solution which has the asymptotics
of $\exp(\alpha_1 z^2)$ in the vicinity of $\varphi+\theta_2/2=0$. But
then, continuing it through the angle of $\pi$ in $\arg(z)=\varphi$,
we would end up in a sector containing the next problematic line
$\varphi+\theta_2/2=\pi$. The corresponding contour in the $y$ plane
would have to rotate through $2\pi$ and the Stokes phenomenon would
cause the asymptotic expansion to now contain also the series with
$\exp(\alpha_2 z^2)$ whose norm would diverge in the second
sector. Thus regardless of the Stokes multipliers, although the
solutions are both entire, none of them is normalisable for $-1\leq x
\leq 1$. One could interpret this as a continuum spectrum with
generalised eigenstates, because to each entire function in the
Bargmann space corresponds a well behaved distribution solution of the
original problem in the $L^2(\mathbb{R})$ space \cite{Bargmann:61::}.

Taking the above into account, we can finally give conditions for the
original entire function to be normalisable. First, for the desired
asymptotics one must have $|\alpha_i|<1/2$ so that the solution starts
as normalisable within some Stokes sector, and second
$\beta_j\in\mathbb{N}$, so that it remains such, as it is continued
throughout the whole complex plane. The second condition is going to
be of the form $\pm\rho-3/4\in\mathbb{N}$
or $\pm\rho-1/4\in\mathbb{N}$ as there are two solutions
of \eqref{lapeq} to choose from, but these can be written together
as $\pm4\rho=2n+1$, $n\in\mathbb{N}$.

\begin{figure}[ht]
  \begin{center}
    \includegraphics[width=\columnwidth]{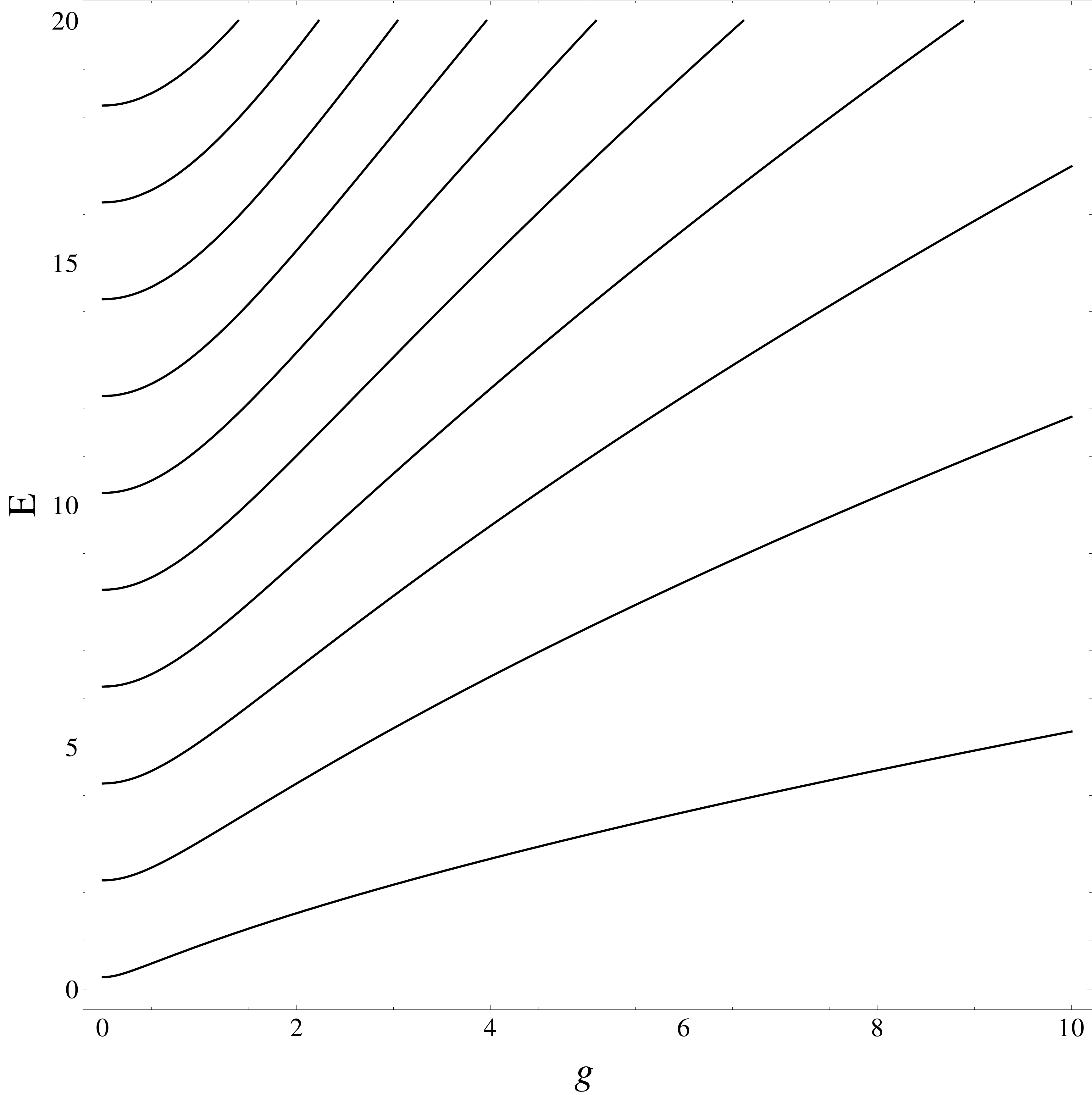}
    \caption{The upper spectrum for $\omega=1$ and $\omega_0=1/2$.}
    \label{fig5}
  \end{center}
\end{figure}

\begin{figure}[ht]
  \begin{center}
    \includegraphics[width=\columnwidth]{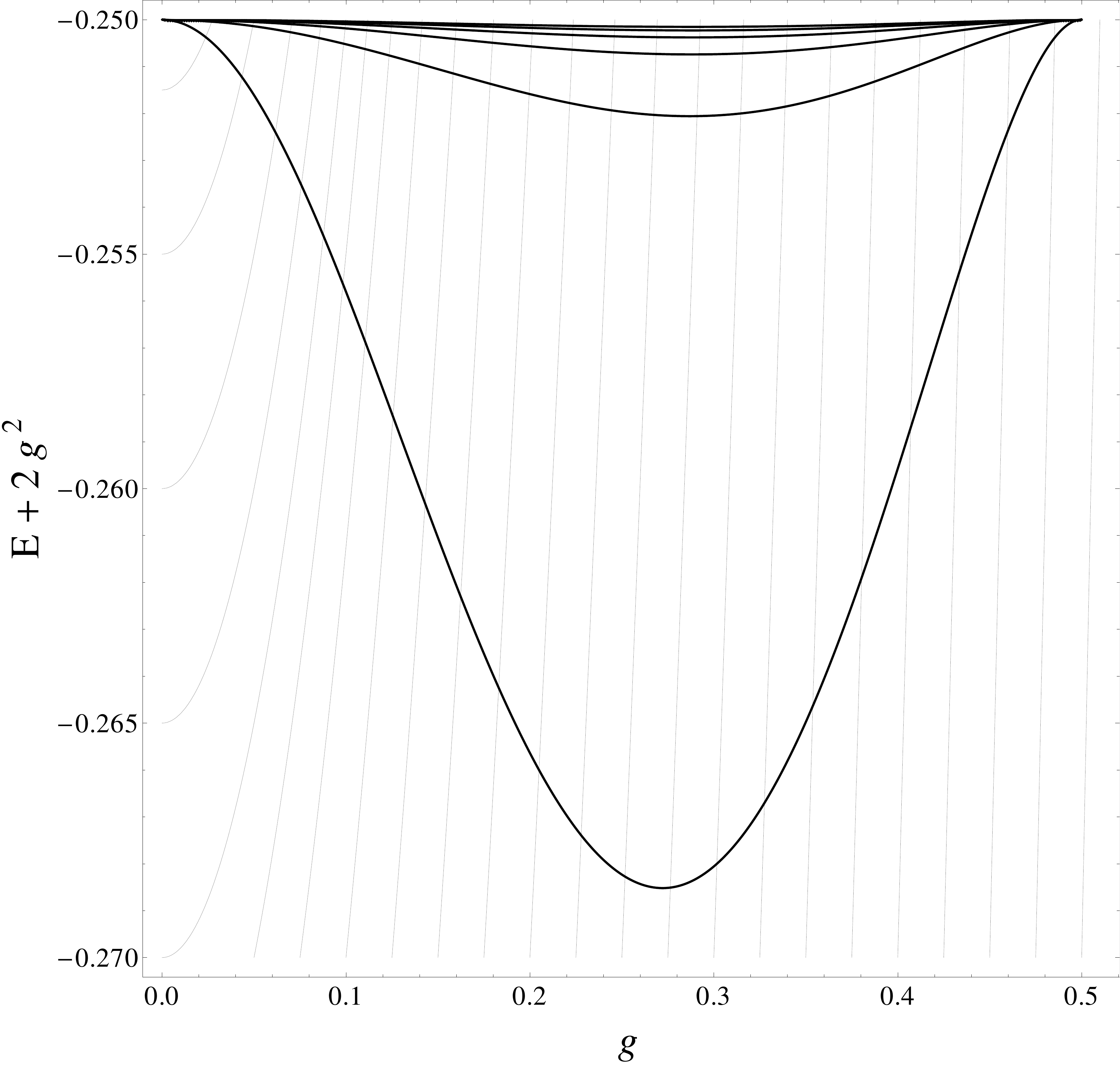}
    \caption{The lower spectrum for $\omega=1$ and
      $\omega_0=1/2$. Thin lines denote constant energy levels.}
    \label{fig6}
  \end{center}
\end{figure}

Returning to the spectral parameter $x$, the normalisation, and
quantisation, conditions read:
\begin{align}
  &\left\{\begin{aligned}
      |\alpha_1| &<\frac12 \iff x>1,\\
      -4\rho&=2n_{+}+1,\quad n_{+}\in\mathbb{N},
    \end{aligned}\right.\\
  &\left\{\begin{aligned}
      |\alpha_2| <\frac12 \iff x<-1,\\
      4\rho=2n_{-}+1,\quad n_{-}\in\mathbb{N}.
    \end{aligned}\right.
\end{align}
or, equivalently,
\begin{equation}
  \mathrm{sgn}(x)\frac{(x-1)\left(\omega(\omega-\omega_0)+g^2(x-1)\right)}
  {\omega^2\sqrt{x^2-1}} = 2n+1,
\end{equation}
for $x^2>1$ and $n\in\mathbb{N}$. The energies of these spectra are presented
in Fig. \ref{fig5} and \ref{fig6}; note that the lower levels become denser as
$E$ gets closer to the
continuous (non-normalisable) region $x^2<1$.

It is evident, that with these
conditions there will always be one characteristic exponent that is a
negative integer. In this case one of the contours $C$ will be
equivalent to a circle and the corresponding Laplace integral can be
immediately evaluated using Cauchy's theorem. It will yield a finite
number of terms so that the solution will be a polynomial in $z$ with
an exponential factor. A change of variables then reveals that each
interval of $x$ has its own sets of eigenstates expressible in terms
of Hermite polynomials, namely
\begin{equation}
  \psi_1(z) = c_1 \exp\left(\frac{-x-\sqrt{x^2-1}}{2}z^2\right)
  H_{n_{-}}\left(\sqrt[4]{x^2-1}\, z\right),
\end{equation}
when $ x<-1$, or
\begin{equation}
  \psi_1(z) = c_2 \exp\left(\frac{-x+\sqrt{x^2-1}}{2}z^2\right)
  H_{n_{+}}\left(\mathrm{i}\sqrt[4]{x^2-1}\, z\right),
\end{equation}
when $x>1$, and in each case the other component is
\begin{equation}
  \psi_2(z) = \frac{\omega}{g(x-1)}\left(z\psi_1(z)+\psi_1'(z)\right).
\end{equation}

\section{Connection with the Whittaker equation}

Let us rewrite system~\eqref{eq:bargmanki} as one equation of the
second order. It reads
\begin{equation}
  \label{eq:3}
  w'' +2 xz w' + \left(z^2+x-m\sqrt{x^2-1}  \right)w=0, 
\end{equation}
where $ w(z)=\psi_1(z)$, $x$ is defined by \eqref{eq:2}, and
\begin{equation}
  \label{eq:4}
  m:= \frac{(x-1)\left[\omega(\omega-\omega_0)+g^2(x-1)\right]}
  {\omega^2\sqrt{x^2-1}}.
\end{equation}
Now, we introduce new independent variable $y$, and change the
depended variable putting $ y=z^2\sqrt{x^2-1}$ and 
\begin{equation}
  \label{eq:5}
  v(y) = y^{1/4}\exp \left[ \frac{xy}{2 \sqrt{x^2-1}} \right]w(z).  
\end{equation}
We obtain the Whittaker equation
\begin{equation}
  \label{eq:6}
  v''(y) - \left( \frac{1}{4}-\frac{\kappa}{y} + \frac{4\mu^2-1}{y^2}
  \right)v(y)=0,
\end{equation}
with parameters
\begin{equation}
  \label{eq:7}
  \kappa = -\frac{1}{4}m, \qquad \mu=\frac{1}{4}.
\end{equation}
According to Proposition~2.5 in~\cite{Morales:99::c}, there exists a basis of solutions of this equation such that two Stokes matrices have the form 
\begin{equation}
\label{eq:8}
S_1=\begin{bmatrix}
1 & \alpha \\
0 & 1
\end{bmatrix}
\quad\text{and}\quad 
S_2=\begin{bmatrix}
1 & 0\\
\beta & 1
\end{bmatrix},
\end{equation}
where $\alpha$ and $\beta$ are some complex numbers. 
Moreover, the Stokes multiplier $\alpha$ vanishes if and only
if either $\kappa+\mu$, or $\kappa-\mu$ is a non-negative half
integer. Similarly, the Stokes multiplier $\beta$ vanishes if and
only if either $-\kappa-\mu$, or $-\kappa+\mu$ is a non-negative half
integer. The reader can check, that in the considered case these
conditions exactly coincide with those given by~\eqref{w+}
and~\eqref{w-}.

Now, it is interesting to observe that for the Whittaker equation if one
of the Stokes multipliers vanishes, then it is solvable in the sense
of differential Galois theory---all its solutions are
Liouvillian. More precisely, the identity component of its
differential Galois group is Abelian, see Corollary~2.1
in~\cite{Morales:99::c}. We have the same  conclusions for
equation~\eqref{eq:3} describing our problem. Thus, if $E$ is an eigenvalue
then:
\begin{enumerate}
\item\label{item:1} A Stokes multiplier vanishes;
\item All solution are Liouvillian;
\item The identity component of its differential Galois group is solvable.  
\end{enumerate}

These observations are very important for application.  If we are able
to prove inverse implications, then we have a strong  tool for
quantisation and finding exactly solvable systems. In fact,
for low order equations there exist effective algorithms for studying
their differential Galois groups and Liouvillian solutions.

Concerning the original equation, we have the full characterisation of its
local
Galois group at infinity. Although in this case it coincides with the full
global group, the conclusions of \cite{Fauvet:10::} suggest, that for systems with
other singular points, the Galois group at infinity alone gives quantisation
conditions. Namely, for energy values which belong to the spectrum, the group of
the corresponding time-independent Schr\"odinger equation is solvable.

\section{Conclusions}
 According to our knowledge, this is the first example
of application of  conditions deduced from
  an analysis of the Stokes phenomenon for a quantum system in the Bargmann
representation.  We only know about application of the finite norm condition for
square-integrable functions to determine spectra for one-dimensional
quantum systems in position representations. Finiteness of this norm
implies some conditions on local solutions containing real 
$\pm\infty$, see e.g. quantisation of quartic oscillator
in \cite{Osherov:11::}, and related problem of Stokes phenomena for
prolate spheroidal wave equation considered
in \cite{Fauvet:10::}. In our considerations, asymptotic solutions
at all sectors, not just real infinity, are involved. It appears that in some
cases these conditions are calculable analytically and then one can obtain 
the spectrum explicitly.

\section{Acknowledgements}

The authors wish to thank M.~Ku\'s for stimulating discussions.  This
research has been supported by grant No.~DEC-2011/02/A/ST1/00208 of
National Science Centre of Poland.


\def\cprime{$'$} \def\cydot{\leavevmode\raise.4ex\hbox{.}} \def\cprime{$'$}
  \def\polhk#1{\setbox0=\hbox{#1}{\ooalign{\hidewidth
  \lower1.5ex\hbox{`}\hidewidth\crcr\unhbox0}}} \def\cprime{$'$}

\end{document}